# Computing the sparse matrix vector product using block-based kernels without zero padding on processors with AVX-512 instructions

Bérenger Bramas and Pavel Kus

Application Group, Max Planck Computing and Data Facility, Garching, Allemagne

## ABSTRACT

The sparse matrix-vector product (SpMV) is a fundamental operation in many scientific applications from various fields. The High Performance Computing (HPC) community has therefore continuously invested a lot of effort to provide an efficient SpMV kernel on modern CPU architectures. Although it has been shown that block-based kernels help to achieve high performance, they are difficult to use in practice because of the zero padding they require. In the current paper, we propose new kernels using the AVX-512 instruction set, which makes it possible to use a blocking scheme without any zero padding in the matrix memory storage. We describe mask-based sparse matrix formats and their corresponding SpMV kernels highly optimized in assembly language. Considering that the optimal blocking size depends on the matrix, we also provide a method to predict the best kernel to be used utilizing a simple interpolation of results from previous executions. We compare the performance of our approach to that of the Intel MKL CSR kernel and the CSR5 open-source package on a set of standard benchmark matrices. We show that we can achieve significant improvements in many cases, both for sequential and for parallel executions. Finally, we provide the corresponding code in an open source library, called SPC5.

Corresponding author
Bérenger Bramas,
berenger.bramas@mpcdf.mpg.de

## INTRODUCTION

The sparse matrix-vector product (SpMV) is an important operation in many applications, which often needs to be performed multiple times in the course of the algorithm. It is often the case that no matter how sophisticated the particular algorithm is, most of the CPU time is spent in matrix-vector product evaluations. The prominent examples are iterative solvers based on Krylov subspaces, such as the popular CG method. Here the solution vector is found after multiple matrix-vector multiplications with the same matrix. Since in many scientific applications a large part of the CPU time is spent in the solution of the resulting linear system and the matrix is stored in a sparse manner, improving the efficiency of the SpMV on modern hardware could potentially leverage the performance of a wide range of codes.

One of the possible approaches towards the improvement of the SpMV is to take advantage of a specific sparsity pattern. For example, the diagonal storage (DIA) (*Saad, 1994*) or jagged diagonal storage (JAD) (*Saad, 2003*) are designed for matrices that are mostly diagonal or band-diagonal. However, unless ensured by the matrix producing method design, it is not straightforward to evaluate in advance when a given matrix's structure is well suited for a specific format. This has been the main motivation to provide general SpMV like block-based schemes, as described in (*Vuduc, 2003*; *Im, Yelick & Vuduc, 2004*; *Vuduc & Moon, 2005*; *Vuduc, Demmel & Yelick, 2005*; *Im & Yelick, 2001*). These types of kernels make it possible to use the SIMD/vectorization capability of the CPU, but they also are required to fill the blocks with zeros to avoid a transformation of the loaded values from the memory before computation. The extra memory usage and the ensuing transfers drastically reduce the effective performance. Moreover, it has been shown that there is no ideal block size that works well for all matrices.

In the current study, we attempt to address the mentioned issues, namely problems caused by zero padding and optimal block selection. The recent AVX-512 instruction set provides the possibility to load fewer values than a vector can contain and to expand them inside the vector (dispatching them in order). This feature allows for fully vectorized block-based kernels without any zero padding in the matrix storage. Additionally, we provide a method to select the block size that is most likely to provide the best performance analyzing the execution times of the previous runs.

The contributions of the study are the following:

- we study block based SpMV without padding with AVX-512,
- we describe an efficient implementation targeting the next generation of Intel's HPC architecture,
- we provide a record-based strategy to choose the most appropriate kernel (using polynomial interpolation in sequential, and linear regression for shared-memory parallel approaches),
- the paper introduces the SPC5 package that includes the source code related to the present study (https://gitlab.mpcdf.mpg.de/bbramas/spc5).

The rest of the paper is organized as follows: 'Background' gives background information related to vectorization and SpMV. We then describe our method in 'Design of Block-based SpMV without Padding' and show its performance on selected matrices of various properties in 'Performance Analysis'. Comparison of various variants of our method using different blocking is provided. In 'Performance Prediction and Optimal Kernel Selection' we propose a systematic approach towards the selection of an optimal kernel based on easily obtained properties of the given matrix. Finally, we draw conclusions in 'Conclusions'.

## BACKGROUND

In this section we recall a few rather well known facts regarding the vectorization in modern processor architectures and about existing sparse matrix vector product implementations.

## Vectorization

The growth of computing power with new generations of processors has been advancing for decades. One of its manifestations, which is of particular interest for scientific computing, is the steady growth of peak performance in terms of the number of floating point operations per second. What has changed recently, however, is the way the hardware manufacturers sustain this growth. With the effective halt in the growth of processor frequencies, most of the increase of computing power is achieved through increasing the number of cores and the ability of each individual core to perform each operation on a vector of certain length using one instruction only. This capability is named vectorization or single instruction on multiple data (SIMD).

The AVX-512 (*Intel, 2016*) is an extension to the *x86* instruction set dedicated to vectorization. It supports a 512 bits vector length, which corresponds to 16 single precision or eight double precision floating point values, and we use the term *VEC_SIZE* to refer to the number of values inside a vector. This instruction set provides new operations such as load/store operations to move data between the main memory and the CPU's registers. One of these new instructions is the *vexpandpd(mask,ptr)*, where *mask* is an unsigned integer of *VEC_SIZE* bits and *ptr* a pointer to an array of single or double precision floating point values. This instruction loads one value for each bit that is set to one in the *mask*, and move them to the corresponding position in the output vector. For example, in double precision, *vexpandpd(10001011b,ptr)* returns a vector equal to *[ptr[0], ptr[1], 0, ptr[2], 0, 0, 0, ptr[3]]*, where *ptr[i]* refers to the values at position *i* in the array of address *ptr*, and considering that the mask is written from right to left. One can see this instruction as a scatter from the main memory to the vector.

## Sparse Matrix Vector Product (SpMV)

If there is any advantage of exploiting the zeros, for example, by saving time or memory, then the matrix should be considered as sparse (*Wilkinson et al., 1971*). However, removing the zeros from the matrix leads to new storage and new computational kernels. While the gain of using a sparse matrix instead of a dense one can be huge in terms of memory occupancy and speed, the effective *Flop* rate of a sparse kernel generally remains low compared to its dense counterpart. In fact, in a sparse matrix storage, we provide a way to know the respective column and row of each non-zero value (NNZ). Therefore, the general SpMV is a bandwidth/memory bound operation because it pays the price of this extra storage and leads to a low ratio of *Flop* to perform against data occupancy. Moreover, the sparsity structure of the matrix makes it difficult to have data reuse or to use vectorization.

The compressed row storage (CRS), also known as the compress sparse row (CSR) storage (*Barrett et al., 1994*), is a well-known storage and is used as a de-facto standard in SpMV studies. Its main idea is to avoid storing individual row indexes for each NNZ value. Instead, it counts the number of values that each row contains. Figure 1 presents an example of the CRS storage. The NNZ values of the original matrix are stored in a *values* array in row major (one row after the other) and in column ascending order. In a secondary array *colidx* we store the column indexes of the NNZ values in the same order.

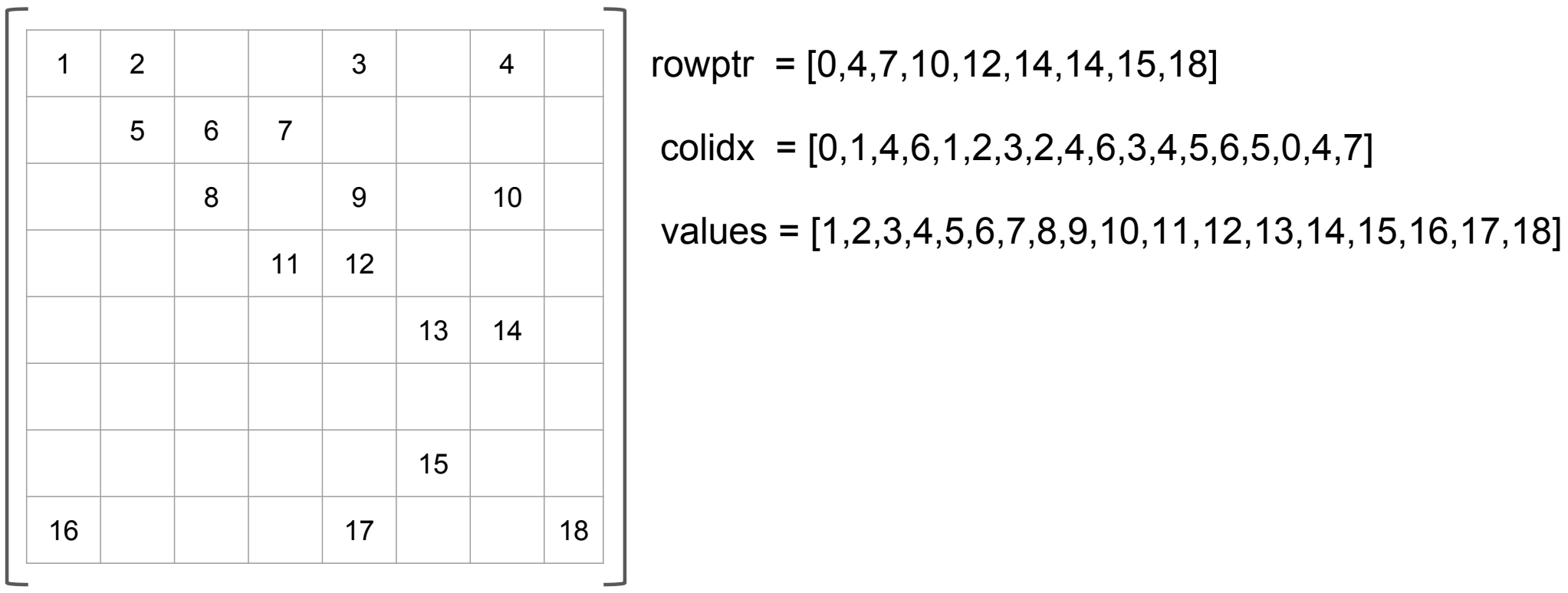

**Figure 1** CSR example.

Full-size DOI: 10.7717/peerjcs.151/fig-1

Finally, *rowptr* contains the positions of the NNZ in *values* for each row: the row *i* has NNZ from index *rowptr[i]* to *rowptr[i+1]-1*. During the computation, the values are read one row after the other, making the access to the result vector linear and potentially unique. Moreover, the input vector is read from left to right at each row computation. The data occupancy is given by $S_{CRS} = N_{NNZ} \times (S_{integer} + S_{float}) + S_{integer} \times (N_{rows} + 1)$, with $N_{rows}$ the number of rows in the original matrix, and $S_{float}$ and $S_{integer}$ the sizes of a floating point value and an integer, respectively. The performance and data locality have been studied in *White III & Sadayappan (1997)*, where the compressed column storage (CCS) variant has also been proposed.

Various papers have shown that there is a need for register blocking, cache blocking, and if possible, multiplication by multiple vectors in *Vuduc (2003)* and *Im (2000)*. In *Toledo (1997)*, the authors introduced the fixed-size block storage (FSB) which is one of the first block-based matrix formats. The key idea is to extract contiguous blocks of values and to process them differently and more efficiently. This makes it possible to take advantage of the contiguous values. The pre-processing of the matrix, that is, the transformation of a matrix from coordinates (COO) or CSR to FSB, is costly but can be beneficial after one or several SpMVs. However, the corresponding SpMV is penalized by the memory accesses because it has to iterate over the vectors several times, canceling the possible memory reuse when all the values are computed together. Subsequently, the block compressed sparse row storage (BCSR) was proposed in *Pinar & Heath (1999)*, and it has been extended with larger blocks of variable dimensions in *Vuduc (2003)* and *Im, Yelick & Vuduc (2004)*. Blocks of values are extracted from the sparse matrix but they had to be filled with zeros to become dense. For these formats, the blocks were aligned (the upper-left corner of the blocks start at a position multiple of the block size). The unaligned block compressed sparse row (UBCSR) has been proposed in *Vuduc & Moon (2005)* where the blocks can start at any row or column. Choosing the block size is not straightforward and as such, some work has been done to provide a mechanism to find it (*Vuduc, Demmel & Yelick, 2005*; *Im & Yelick, 2001*).

The main drawback of the compressed sparse matrix format is that the data locality is not preserved and it is thus more difficult to vectorize the operations. A possible attempt to solve this problem is by combining a sparse and dense approach: a certain block size is selected and the matrix is covered by blocks of this size so that all non-zero elements of the matrix belong to some block. The positions of the blocks are then stored in sparse fashion (using row pointers and column indices), while each block is stored as dense, effectively by storing all elements belonging to the block explicitly, including zeros. This leads to padding the non-zero values in the *values* array by zeros and thus increases memory requirements. The immense padding implied by their design led to the failure to adopt these methods in real-life calculations.

The authors from *Yzelman (2015)* show how to use gather/scatter instructions to compute block-based SpMV. However, the proposed method still fill the blocks with zeros in the matrix storage to ensure that blocks have values of a fixed size. Moreover, they use arrays of integers, needed by the scatter/gather operations, which adds important memory occupancy to the resulting storage. The author also describe the bit-based methods as not efficient in general, but we show in the current study that approaches are now efficient. The proposed mechanism from *Buluc et al. (2011)* is very similar to our work. The authors design a SpMV using bit-masks. However, they focus on symmetric matrices and build their work on top of SSE, which requires several instructions to do what can be done in a single now. In *Kannan (2013)*, the authors use bitmasks to represent the positions of the NNZ inside the blocks as we do here. However, they use additional integers to represent the position of the blocks in the matrix, while we partially avoid aligning the block vertically. In addition, they fail to develop a highly-tuned and optimized version of their kernel in order to remain portable. Consequently, they do not use vectorization explicitly and their implementation is not parallel.

More recent work has been done, pushed by the research on GPUs and the availability of manycore architectures. In *Liu et al. (2013)*, the authors extend the ELLAPACK format that became popular due to its high performance on GPUs, and adapt it to the Intel KNC. They provide some metrics to estimate if the computation of a matrix is likely to be memory or computation bounded. They conclude that block-based schemes are not expected to be efficient because the average number of NNZ per block can be low. As we will show in the current study, this is only partially true because block-based approaches require less transformation of the input matrix and in extreme cases it is possible to use the block mask to avoid useless memory load. The authors of *Liu et al. (2013)* also propose an auto-tuning mechanism to balance the work between threads, and their approach appears efficient on the KNC. The authors of *Kreutzer et al. (2014)* define the SELL-C- $\sigma$ format as a variant of Sliced ELLPACK. The key idea of their proposal is to provide a single matrix format that can be used on all common HPC architectures including regular CPUs, manycore CPUs, and GPUs. We consider that focusing on CPUs only could lead to better specific matrix storage. In *Liu & Vinter (2015)* the authors also target CPUs and GPUs and introduce a new matrix format, called CSR5. A corresponding source code is freely available online. We include their code in our performance benchmark.

### Matrix permutation/reordering

Permutation of the rows and/or columns of a sparse matrix can improve the memory access pattern and the storage. A well-known technique called Cuthill–McKee from *Cuthill & McKee (1969)* tries to make a matrix bandwidth by applying a breadth-first algorithm on a graph which represents the matrix structure such that the resulting matrices have good properties for LU decomposition. However, the aim of this algorithm is not to improve the SpMV performance even though the generated matrices may have better data locality.

In *Pinar & Heath (1999)*, a method is proposed to have specifically more contiguous values in rows or columns. The idea is to create a graph from a matrix where each column is a vertex and by connecting all the vertices with weighted edges. The weights come from different formulations, but they represent the interest of putting two columns contiguously. Then, a permutation is found by solving the traveling salesman problem (TSP) to obtain a path that goes through all the nodes but only once and that minimizes the cost of the total weight of the path. Therefore, a path in the graph represents a permutation; when we add a node to a path, it means that we aggregate a column to a matrix in construction. The method has been updated in *Vuduc & Moon (2005)*, *Pichel et al. (2005)* and *Bramas (2016)* with different formulas.

The permutation of the matrices has been left aside from the current study but as in most other approaches, any improvement to the shape of the matrix will certainly improve the efficiency of our kernels by reducing the number of blocks.

## DESIGN OF BLOCK-BASED SpMV WITHOUT PADDING

In this section, we will elaborate on an alternative approach to the existing block-based storage by exploiting the mask features from the AVX-512 instruction set. Instead of padding the nonzero values with zeros to fill the whole blocks, we store an array of masks where each mask corresponds to one block and describes how many non-zeros there are and on which positions within the block. We also describe the corresponding SpMV kernels and discuss their optimization and parallelization.

### Block-based storage without zero padding

We refer to $\beta(r,c)$ as the matrix storage that has blocks of size $r \times c$, where $r$ is the number of rows and $c$ the number of columns. Figure 2 shows the same matrix as in Fig. 1 in the formats $\beta(1,4)$ and $\beta(2,2)$, see Figs. 2A and 2B, respectively. In our approach, the blocks are row-aligned i.e., the upper left corner of a block starts at a row of index multiple of $r$, but at any column. To describe the sparse matrix, we need four different arrays as shown in the figures. In the *values* array, we store the NNZ values in block order and in row major inside the blocks. This array remains unchanged compared to the CSR format if we have one row per block ($r = 1$). The *block_colidx* array contains the column indexes of the upper left values of each block. We store in *block_rowptr* the number of blocks per row interval ($r$ consecutive rows). Finally, the *block_masks* array provides one mask of $r \times c$ bits per block to describe the sparsity structure inside each block.

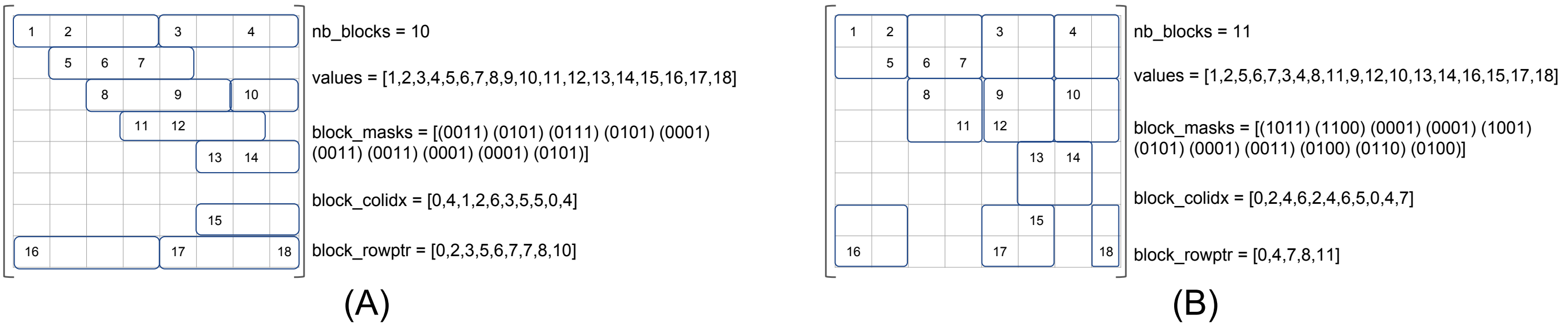

**Figure 2 SPC5 format examples.** The masks are written in conventional order (greater/right, lower/left). (A) SPC5 BCSR Example for $\beta(1,4)$. The *values* array is unchanged compared to the CSR storage. (B) SPC5 BCSR Example for $\beta(2,2)$.
Full-size DOI: 10.7717/peerjcs.151/fig-2

We note $N_{blocks}(r,c)$ the number of blocks of size $r \times c$ obtained from a given matrix. The average number of NNZ per block is then $Avg(r,c) = N_{NNZ}/N_{blocks}(r,c)$. The memory occupancy is given by (in bytes) :

$$\begin{aligned} O(r,c) &= O_{values}(r,c) + O_{block_colidx}(r,c) + O_{block_rowptr}(r,c) + O_{block_masks}(r,c) \\ O_{values}(r,c) &= N_{NNZ} \times S_{float} \\ O_{block_rowptr}(r,c) &\approx \frac{N_{rows}}{r} \times S_{integer} \\ O_{block_colidx}(r,c) &= N_{blocks}(r,c) \times S_{integer} \\ O_{block_masks}(r,c) &= \frac{N_{blocks}(r,c) \times r \times c}{8}, \end{aligned} \tag{1}$$

with $S_{float}$ and $S_{integer}$ the sizes of a floating point value and an integer, respectively. After putting two terms together and substituting for $N_{blocks}(r,c) = N_{NNZ}/Avg(r,c)$ we get the total occupancy

$$O(r,c) = N_{NNZ} \times S_{float} + N_{rows} \times \frac{S_{integer}}{r} + N_{NNZ} \times \frac{8 \times S_{integer} + r \times c}{8 \times Avg(r,c)}. \tag{2}$$

Let us now compare with the memory occupancy of the CSR format, which is

$$O_{CSR} = N_{NNZ} \times S_{float} + N_{rows} \times S_{integer} + N_{NNZ} \times S_{integer}. \tag{3}$$

The first term is the same for both storages. This is thanks to the fact that we do not use zero padding in the values array, as it has been discussed before. The second term is only relevant for the very sparse matrices: otherwise $N_{NNZ} \gg N_{rows}$) is clearly either the same (if $r = 1$) or smaller for our storage whenever $r > 1$. The last term is smaller for our storage if

$$Avg(r,c) > 1 + \frac{r \times c}{8 \times S_{integer}}, \tag{4}$$

which should be usually true unless the blocks are very poorly filled. If we consider the usual size of the integer $S_{integer} = 4$, we need average filling of at least $1 + \frac{1}{4}$ for $\beta(1,8)$, $1 + \frac{1}{2}$ for $\beta(2,8)$ and $\beta(4,4)$, and 2 for $\beta(4,8)$ and $\beta(8,4)$, respectively.

Memory occupancy is an important measure, because the SpMV is usually a memory-bound operation and therefore, reducing the total amount of memory to perform the same number of floating point operations (Flop) is expected to be beneficial by shifting to a more computational-bound kernel. Indeed, using blocks helps to remove the usual SpMV limits in terms of a poor data reuse of the NNZ values and non-contiguous accesses on the dense vectors.

## SpMV kernels for $\beta(r,c)$ storages

We provide the SpMV kernel that works for any block size in Algorithm 1. The computation iterates over the rows with a step $r$ since the blocks are $r$-aligned. Then, it iterates over all the blocks inside a row interval, from left to right, starting from the block of index *mat.block_rowptr[idxRow/r]* and ending before the block of index *mat.block_rowptr[idxRow/r+1]* (line 7). The column index for each block is given by *mat.block_colidx* and stored in *idxCol* (line 8). To access the values that correspond to a block, we must use a dedicated variable *idxVal*, which is incremented by the number of bits set to 1 in the masks. The scalar algorithm relies on an inner loop of index $k$ to iterate over the bits of the masks. This loop can be vectorized in AVX-512 using the *vexpand* instruction such that the arithmetic operations between $y$, $x$ and *mat.values* are vectorized.

In Algorithm 1, all the blocks from a matrix are computed similarly regardless whether their masks contain all ones or all zeros. This can become unfavorable in the case of extremely sparse matrices: most of the blocks will then contain a single value and only one bit of the mask will be equal to 1. This implies two possible overheads; first we load the values from the matrix using the *vexpand* instruction instead of a scalar move, and second, we load a full vector from $x$ and use vector-based arithmetic instruction. This is why we propose an alternative approach shown in Algorithm 2 for $\beta(1, VEC_SIZE)$. The idea is to use two separate inner loops to proceed differently on the blocks depending on whether they contain only one value or more than one value. However, having a test inside the inner loop could kill the instruction pipeline and the speculation from the CPU. Instead, we use two separate inner loops and jump from one loop to the other one when needed. Indeed, using *goto* command might seem strange, but it is justified by performance considerations and, since the algorithm is implemented in assembly, it is a rather natural choice. Regarding the performance, we expect that for most matrices the algorithm would stay in one of the modes (scalar or vector) for several blocks and thus, the CPU is more likely to predict to stay inside the loop, avoiding the performance penalty. Therefore, the maximum overhead is met if the blocks' kinds alternate such that the algorithm jumps from one loop to the other at each block. Still, this approach can be significantly beneficial in terms of data transfer especially if the structure of the matrix is chaotic. In the following, we refer the kernels that use such a mechanism as $\beta(x,y)$ *test* to indicate that they use a test inside the computational loop.

**ALGORITHM 1:** SpMV for a matrix *mat* in format $\beta(r,c)$. The lines in blue are to compute in scalar and have to be replaced by the line in green to have the vectorized equivalent.

```
Input: x : vector to multiply with the matrix. mat : a matrix in the block format β(r,c). r, c : the size of the blocks.
Output: y : the result of the product.
function spmv(x, mat, r, c, y)
    // Index to access the array's values
    idxVal ← 0
    for idxRow ← 0 to mat.numberOfRows-1 inc by r do
        sum[r] ← init_scalar_array(r, 0)
        sum[r] ← init_simd_array(r, 0)
        for idxBlock ← mat.block_rowptr[idxRow/r] to mat.block_rowptr[idxRow/r+1]-1 do
            idxCol ← mat.block_colidx[idxBlock]
            for idxRowBlock ← 0 to r do
                valMask ← mat.block_masks[idxBlock × r + idxRowBlock]
                // The next loop can be vectorized with vexpand
                for k ← 0 to c do
                    if bit_shift(1 , k) BIT_AND valMask then
                        sum[idxRowBlock] += x[idxCol+k] * mat.values[idxVal]
                        idxVal += 1
                    end
                end
                // To replace the k-loop
                sum[idxRowBlock] += simd_load(x[idxCol]) * simd_vexpand(mat.values[idxVal], valMask)
            end
        end
        for idxRowBlock ← 0 to r do
            y[ridxRowBlock] += sum[r]
            y[ridxRowBlock] += simd_hsum(sum[r])
        end
    end
```

## Relation between matrix shape and number of blocks

The number of blocks and the average number of values inside the blocks for a particular block size provide only limited information about the structure of a given matrix. For example, having a high filling of the blocks means that locally, the non-zero values are close to each other, but it says nothing of the possible memory jump in the $x$ array from one block to the next one. If, however, this information is known for several block sizes, more can be deduced about the global structure of the given matrix. Having small blocks largely filled and large blocks poorly filled suggests that there is hyper-concentration, but still, gaps between the blocks. On the other hand, the opposite would mean that the NNZs are not far from each other, but not close enough to fill small blocks. In the present study, we try to predict the performance of our different kernels (using different block sizes) using the average number of values per block with the objective of selecting the best kernels, before converting a matrix into the format required by our algorithm.

## Optimized kernel implementation

In 'SpMV Kernels for $\beta(r,c)$ storages', we described generic kernels, which can be used with any block sizes. For the most useful block sizes, which we consider to be $\beta(1,8)$, $\beta(2,4)$, $\beta(2,8)$, $\beta(4,4)$, $\beta(4,8)$ and $\beta(8,4)$, we decided to develop highly optimized routines in assembly to further reduce the run-time. In this section, we describe some of the technical considerations leading to the optimized kernels speed-up.

**ALGORITHM 2:** Scalar SpMV for a matrix *mat* in format $\beta(1, VEC_SIZE)$ with test (vectorized in the second *idxBlock* loop).

```
Input: x : vector to multiply with the matrix. mat : a matrix in the block format.
Output: y : the result of the product.
function spmv_scalar_1_VECSIZE(x, mat, y)
    // Index to access to the values array
    idxVal ← 0
    for idxRow ← 0 to mat.numberOfRows-1 inc by 1 do
        sum_scalar ← 0
        sum_vec ← simd_set(0)
        // Loop for mask equal to 1
        for idxBlock ← mat.block_rowptr[idxRow] to mat.block_rowptr[idxRow]-1 do
            idxCol ← mat.block_colidx[idxBlock]
            valMask ← mat.block_masks[idxBlock]
            if valMask not equal to 1 then
                goto loop-not-1
            end
            label loop-for-1:
            sum_scalar += x[idxCol] * mat.values[idxVal]
            idxVal += 1
        end
        goto end-of-loop
        for idxBlock ← mat.block_rowptr[idxRow] to mat.block_rowptr[idxRow]-1 do
            idxCol ← mat.block_colidx[idxBlock]
            valMask ← mat.block_masks[idxBlock]
            if valMask equal to 1 then
                goto loop-for-1
            end
            label loop-not-1:
            vec_sum += simd_load(x[idxCol]) * simd_vexpand(mat.values[idxVal], valMask)
            idxVal += pop_count(valMask)
        end
        label end-of-loop:
        y[idxRowBlock+idxRow] += sum_scalar + simd_hsum(sum_vec);
    end
```

In AVX-512, *VEC_SIZE* is equal to 8, so the formats that have blocks with four columns load only half a vector from $x$ into the registers. In addition, the *vexpand* instruction loads values for two consecutive rows of the block. Consequently, we have to decide in the implementation between expanding the half vector from $x$ into a full AVX-512 register or splitting the values vector into two AVX-2 registers. We have made the choice of the second option.

We have decided to implement our kernels in assembly language. By employing register-oriented programming, we intend to reduce the number of instructions (compared to a *C/C++* compiled code) and to minimize the access to the cache. Thus, we achieve some non-temporal usage of all the arrays related to the matrix. Moreover, we are able to apply software pipelining techniques, even though it is difficult to figure out how the hardware is helped by this strategy. We have compared our implementation to an intrinsic-based *C++* equivalent and got up to 10% difference (comparison not included in the current study). We provide a simplified source code of the $\beta(1,8)$ kernel in Code 1.

## Parallelization

We parallelize our kernels with a static workload division among OpenMP threads. Our objective is to have approximately the same number of blocks per thread, which

```
extern "C" void core_SPC5_1rVc_Spmv_asm_double(const long int nbRows, const int* rowsSizes,
                                               const unsigned char* headers, const double* values,
                                               const double* x, double* y);

// (nbRows rdi , rowsSizes rsi , headers rdx , values rcx, x r8, y r9 )
// save some registers
(commented out) ...
xorq %r12, %r12;  // valIdx = 0
// if no rows in the matrix, jump to end
test %rdi, %rdi;
jz   compute_Spmv512_avx_asm_double_out_exp;
    xorq %r10, %r10;  // idxRow/r10 = 0
    compute_Spmv512_avx_asm_double_loop_exp:

    movslq 4(%rsi,%r10,4), %r11;   // rowsSizes[idxRow+1]
    subl   0(%rsi,%r10,4), %r11d;  // nbBlocks = rowsSizes[idxRow+1]-rowsSizes[idxRow]
    // if no blocks for this row, jump to next interval
    jz compute_Spmv512_avx_asm_double_outrow_exp;

    vpxorq %zmm0,%zmm0,%zmm0;        // %zmm0 sum = 0
    compute_Spmv512_avx_asm_double_inrow_exp:
            movslq 0(%rdx), %r13;  // colIdx = *rdx or *headers
            movzbl 4(%rdx), %r14d; // mask = *(rdx+4) or *(headers+4)
            kmovw  %r14d,   %k1;    // mask
            vexpandpd   (%rcx,%r12,8), %zmm1{%k1}{z}; // values (only some of them)
            vfmadd231pd (%r8,%r13,8),  %zmm1, %zmm0;  // mul add to sum

            popcntw %r14w, %r14w; // count the number of bits in the mask
            addq    %r14,  %r12;  // valIdx += number of bits(mask)
            addq    $5,    %rdx;  // headers += 1 * int + 1 mask
            subq    $1,    %r11;  // nbBlocks -=1, if equal zero go to end of interval
    jnz compute_Spmv512_avx_asm_double_inrow_exp;

    compute_Spmv512_avx_asm_double_inrow_exp_stop:
    // Horizontal sum from ymm0 to xmm0
    (commented out) ...
    // add to y, r9[r10] => y[idxRow]
    vaddsd  (%r9,%r10,8), %xmm0, %xmm0;
    vmovsd  %xmm0,        (%r9,%r10,8);
    compute_Spmv512_avx_asm_double_outrow_exp:

    addq $1,   %r10; // idxRow += 1
    cmp  %rdi, %r10; // idxRow == nbRows
    jne compute_Spmv512_avx_asm_double_loop_exp; // if nbBlocks != 0 go to beginning
compute_Spmv512_avx_asm_double_out_exp:
```

Code 1: $\beta(1,8)$ kernel in assembly language.

is $N_{b/t} = N_{blocks}(r,c)/N_{threads}$ in an ideal case, but without distributing one row to multiple threads. We create the row intervals by iterating over the blocks in rows order with a step $r$ and by deciding whether all of the blocks inside the next $r$ rows should be added to the current interval. We add the next $r$ rows if the following test is true: $absolute_value((thread_id+1) \times N_{b/t} - N_{blocks}(r,c)[row_idx])$ is lower than $absolute_value((thread_id+1) \times N_{b/t} - N_{blocks}(r,c)[row_idx+1])$, where $thread_id$ is the index of the current interval that will be assigned to the thread of the same id for computation. We obtain one-row interval per thread, and we pre-allocate a working vector of the same size.

Once all threads finish their respective parts of the calculation, all the partial results are merged into the final vector of size *Dim*. This operation is done without synchronization between the threads since there is no overlap between the rows assigned to each of them. Thus, it is possible for each thread to copy its working vector into the global one directly. Therefore, even in the case of slight work-load imbalance, after a thread has finished, it does not wait for the others but starts to directly add its results.

We attempt to reduce the NUMA effects by splitting the matrix's arrays *values*, *block_colidx*, *block_rowptr* and *block_masks* to allocate sub-arrays for each thread in the memory node that corresponds to the core where it is pinned. There are, however, conceptual disadvantages to this approach since it duplicates the matrix, at least temporarily, in memory during the copy and it ties the data structure and memory distribution to the number of threads. The vectors $x$ and $y$ are still allocated by the master thread, and $x$ is accessed during the computation while $y$ is only accessed during the merge.

# PERFORMANCE ANALYSIS

In the previous section, a family of sparse matrix formats and their corresponding SpMV kernels has been described. In this section, we show their performance using selected matrices from the SuiteSparse Matrix Collection, formerly known as the University of Florida Sparse Matrix Collection (*Davis & Hu, 2011*). We describe our methodology and provide comparisons between our kernels of different block sizes, and also comparisons with MKL and CSR5 libraries. The performance comparisons are presented for serial ('Sequential SpMV performance') and parallel ('Parallel SpMV performance') versions of algorithms, respectively.

## Hardware/software

We used a compute node with two Intel Xeon Platinum 8170 (Skylake) CPUs at *2.10 GHz* and 26 cores each, with caches of sizes *32K*, *1024K* and *36608K*, and the GNU compiler 6.3. We bind the memory allocation using *numaclt –localalloc*, and we bind the processes by setting *OMP_PROC_BIND=true*. As references, we use Intel MKL 2017.2 and the CSR5 package taken from the bhSPARSE repository accessed on the 11th of September 2017 (https://github.com/bhSPARSE/Benchmark_SpMV_using_CSR5). We obtain a floating point operation per second (FLOPS) measure using the formula $2 \times N_{NNZ}/T$, where $T$ is the execution time in seconds. The execution time is measured as an average of 16 consecutive runs without accessing the matrix before the first run.

## Test matrices

We selected matrices that were used in the study (*Ye, Calvin & Petiton, 2014*) and added few more to obtain a diverse set. The matrices labeled Set-A, see Table 1, are used in the computation benchmark. Their execution times are also used in our prediction system, which will be introduced in 'Performance Prediction and Optimal Kernel Selection'.

The user of our library could create the matrix directly using one of our block-based schemes even though it is more likely to be impossible in many cases due to incompatibility with the other parts of the application. If he could, he would be able to choose the most appropriate block size depending on the expected matrix structure. Otherwise, the user will need to convert the matrix from a standard CSR format to one of our formats. The time taken to convert any of the matrices form the Set-A from the CSR format to one of ours is around twice the time of a single SpMV in sequential. For example, for the *atmosmodd* and *bone010*, the conversion takes approximately 0.4 and 4 s, and the sequential execution around 0.2 and 1.5 s, respectively. In a typical scenario, however, many matrix-vector multiplications with the same matrix are performed, so the cost of the matrix conversion can be covered. When the original format is not the CSR, the user will have to convert first to the CSR and then to one of our formats, or to create an own conversion method.

**Table 1** Matrix set for computation and performance analysis. Refered to as Set-A.

| Name | Dim | $N_{NNZ}$ | $\frac{N_{NNZ}}{N_{rows}}$ | $\frac{N_{NNZ}}{N_{blocks(1,8)}}$ | $\frac{N_{NNZ}}{N_{blocks(2,4)}}$ | $\frac{N_{NNZ}}{N_{blocks(2,8)}}$ | $\frac{N_{NNZ}}{N_{blocks(4,4)}}$ | $\frac{N_{NNZ}}{N_{blocks(4,8)}}$ | $\frac{N_{NNZ}}{N_{blocks(8,4)}}$ |
|---|---|---|---|---|---|---|---|---|---|
| atmosmodd | 1,270,432 | 8,814,880 | 6 | 1.4 (18%) | 2.8 (35%) | 2.8 (18%) | 4.7 (29%) | 5.6 (18%) | 5.1 (16%) |
| Ga19As19H42 | 133,123 | 8,884,839 | 66 | 2.4 (30%) | 3.7 (46%) | 4.6 (29%) | 6.6 (41%) | 8.4 (26%) | 7.7 (24%) |
| mip1 | 66,463 | 10,352,819 | 155 | 6.5 (81%) | 7.1 (89%) | 13 (81%) | 14 (88%) | 25 (78%) | 24 (75%) |
| rajat31 | 4,690,002 | 20,316,253 | 4 | 1.4 (18%) | 1.9 (24%) | 1.9 (12%) | 2.1 (13%) | 2.3 (7%) | 2.2 (7%) |
| bone010 | 986,703 | 71,666,325 | 72 | 4.6 (58%) | 5.9 (74%) | 9 (56%) | 11 (69%) | 17 (53%) | 16 (50%) |
| HV15R | 2,017,169 | 283,073,458 | 140 | 5.4 (68%) | 5.7 (71%) | 10 (63%) | 9.7 (61%) | 18 (56%) | 15 (47%) |
| mixtank new | 29,957 | 1,995,041 | 66 | 2.5 (31%) | 3 (38%) | 3.9 (24%) | 3.8 (24%) | 5.5 (17%) | 4.9 (15%) |
| Si41Ge41H72 | 185,639 | 15,011,265 | 80 | 2.6 (33%) | 3.9 (49%) | 5 (31%) | 6.8 (43%) | 9 (28%) | 8.2 (26%) |
| cage15 | 5,154,859 | 99,199,551 | 19 | 1.2 (15%) | 2 (25%) | 2.1 (13%) | 3.1 (19%) | 3.6 (11%) | 3.4 (11%) |
| in-2004 | 1,382,908 | 16,917,053 | 12 | 3.8 (48%) | 4.4 (55%) | 6.2 (39%) | 6.7 (42%) | 9.6 (30%) | 9.6 (30%) |
| nd6k | 18,000 | 6,897,316 | 383 | 6.5 (81%) | 6.6 (83%) | 12 (75%) | 12 (75%) | 23 (72%) | 22 (69%) |
| Si87H76 | 240,369 | 10,661,631 | 44 | 1.8 (23%) | 3 (38%) | 3.4 (21%) | 5.5 (34%) | 6.5 (20%) | 6.1 (19%) |
| circuit5M | 5,558,326 | 59,524,291 | 10 | 2 (25%) | 3.3 (41%) | 3.7 (23%) | 5.5 (34%) | 6.7 (21%) | 6.7 (21%) |
| indochina-2004 | 7,414,866 | 194,109,311 | 26 | 4.6 (58%) | 5.1 (64%) | 7.7 (48%) | 8.3 (52%) | 12 (38%) | 13 (41%) |
| ns3Da | 20,414 | 1,679,599 | 82 | 1.2 (15%) | 1.2 (15%) | 1.3 (8%) | 1.4 (9%) | 1.5 (5%) | 1.5 (5%) |
| CO | 221,119 | 7,666,057 | 34 | 1.5 (19%) | 2.6 (33%) | 2.9 (18%) | 5.1 (32%) | 5.7 (18%) | 5.5 (17%) |
| kron g500-logn21 | 2,097,152 | 182,082,942 | 86 | 1 (13%) | 1 (13%) | 1 (6%) | 1 (6%) | 1 (3%) | 1 (3%) |
| pdb1HYS | 36,417 | 4,344,765 | 119 | 6.2 (78%) | 6.6 (83%) | 12 (75%) | 12 (75%) | 20 (63%) | 20 (63%) |
| torso1 | 116,158 | 8,516,500 | 73 | 6.5 (81%) | 7.5 (94%) | 13 (81%) | 13 (81%) | 25 (78%) | 21 (66%) |
| crankseg 2 | 63,838 | 14,148,858 | 221 | 5.3 (66%) | 6 (75%) | 9.5 (59%) | 9.7 (61%) | 16 (50%) | 15 (47%) |
| ldoor | 952,203 | 46,522,475 | 48 | 7 (88%) | 6.4 (80%) | 13 (81%) | 11 (69%) | 21 (66%) | 17 (53%) |
| pwtk | 217,918 | 11,634,424 | 53 | 6 (75%) | 6.7 (84%) | 12 (75%) | 13 (81%) | 23 (72%) | 21 (66%) |
| Dense-8000 | 8,000 | 64,000,000 | 8,000 | 8 (100%) | 8 (100%) | 16 (100%) | 16 (100%) | 32 (100%) | 32 (100%) |

When the conversion is required, it is worth noting that it is easiest to convert to $\beta(1,8)$ format, since each block is part of a single row and thus the *values* array remains intact and only the index arrays have to be altered. On the other hand, the $\beta(1,8)$ kernel might not be the best performing one, so the optimal decision has to be taken depending on a particular setting. The detailed performance analysis of matrix conversions is not included in the current study.

## Sequential SpMV performance

Figure 3 shows the performance of our kernels in sequential for the matrices from Set-A. We can see significant speed-up in most cases, often up to 50%. We observe that we obtain a speedup even with the $\beta(1,8)$ formats compared to the CSR and CSR5 in many cases (and as we mentioned in the previous section, this is of particular importance due to the ease of conversion from CSR format). However, for all the matrices that have an average number

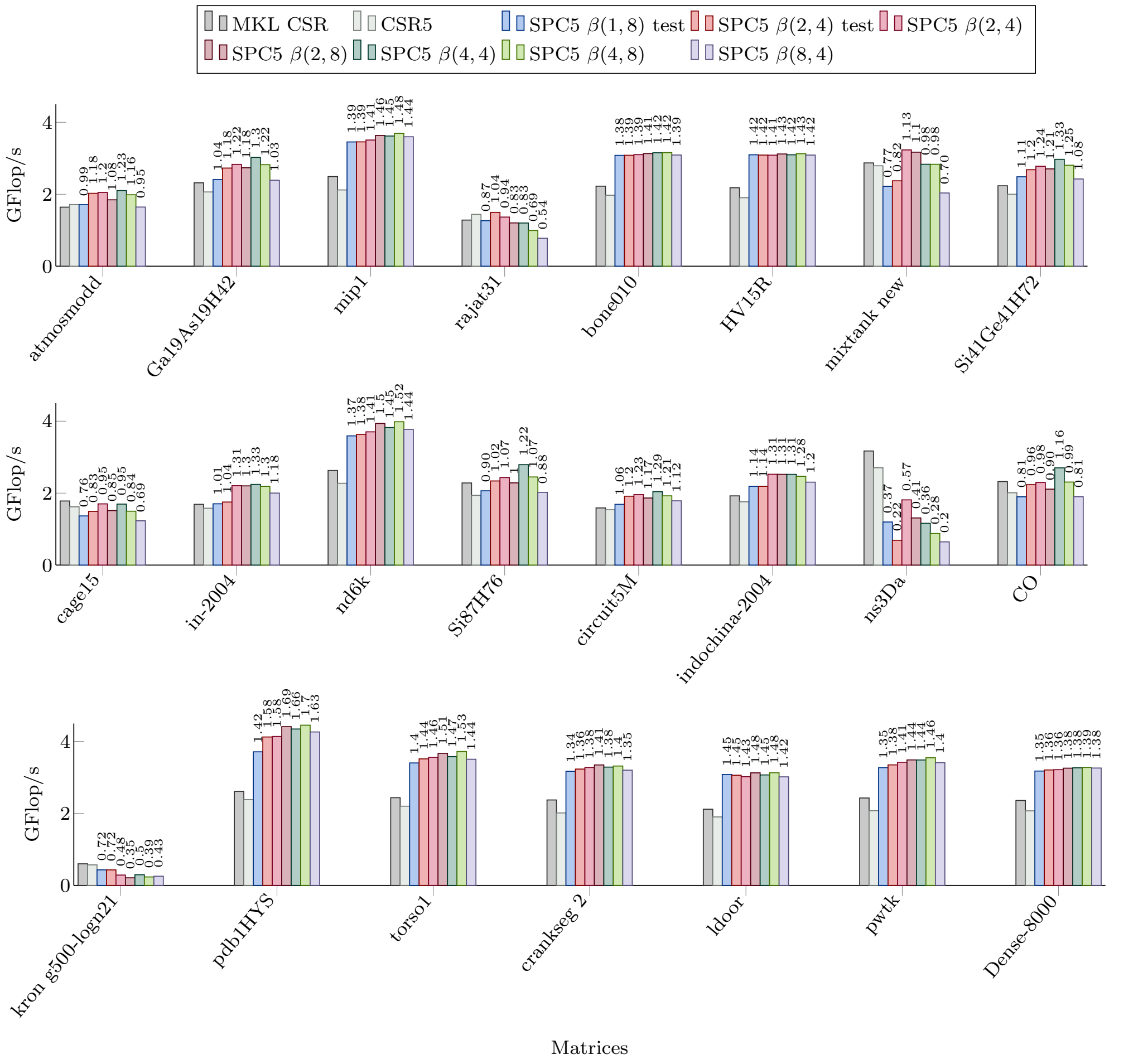

**Figure 3 Performance in Giga Flop per second for sequential computation in double precision for the MKL CSR, the CSR5 and our SPC5 kernels.** Speedup of SPC5 against the better of MKL CSR and CSR5 is shown above the bars.

Full-size DOI: 10.7717/peerjcs.151/fig-3

of non-zero values per block of corresponding storage below 2, the $\beta(1,8)$ is more likely to be slower than the CSR (with some exceptions, e.g., for the *mixtank new* the average is 2.5 and $\beta(1,8)$ is still slower). For the other blocks sizes, we obtain a similar behavior, i.e., if there are insufficient values per block, the performance decreases. The worse case is for the *ns3Da* and *kron g500-logn21* matrices, and we see from Table 1 that the blocks remain unfilled for all the considered storages. That hurts the performance of our block-based kernels, since 8 or 4 values still have to be loaded from $x$, even though only one value is useful, and large width arithmetic operations (multiplication and addition of vectors) still have to be called.

For the $\beta(2,4)$ storage, the test-based scheme provides a speedup only for *rajat31*. For other formats, we see that the performance is mainly a matter of the average number of values per block (so related to the matrix sparsity pattern). In fact, if we look to the *Dense-8000* matrix where all the blocks are completely filled, the performance is not very different from one kernel to the other. For instance, there are more values per block in the $\beta(8,4)$ then in the $\beta(4,8)$ storage only for *indochina-2004* matrix, which means that it did not help to capture 8 rows per block instead of 8 columns.

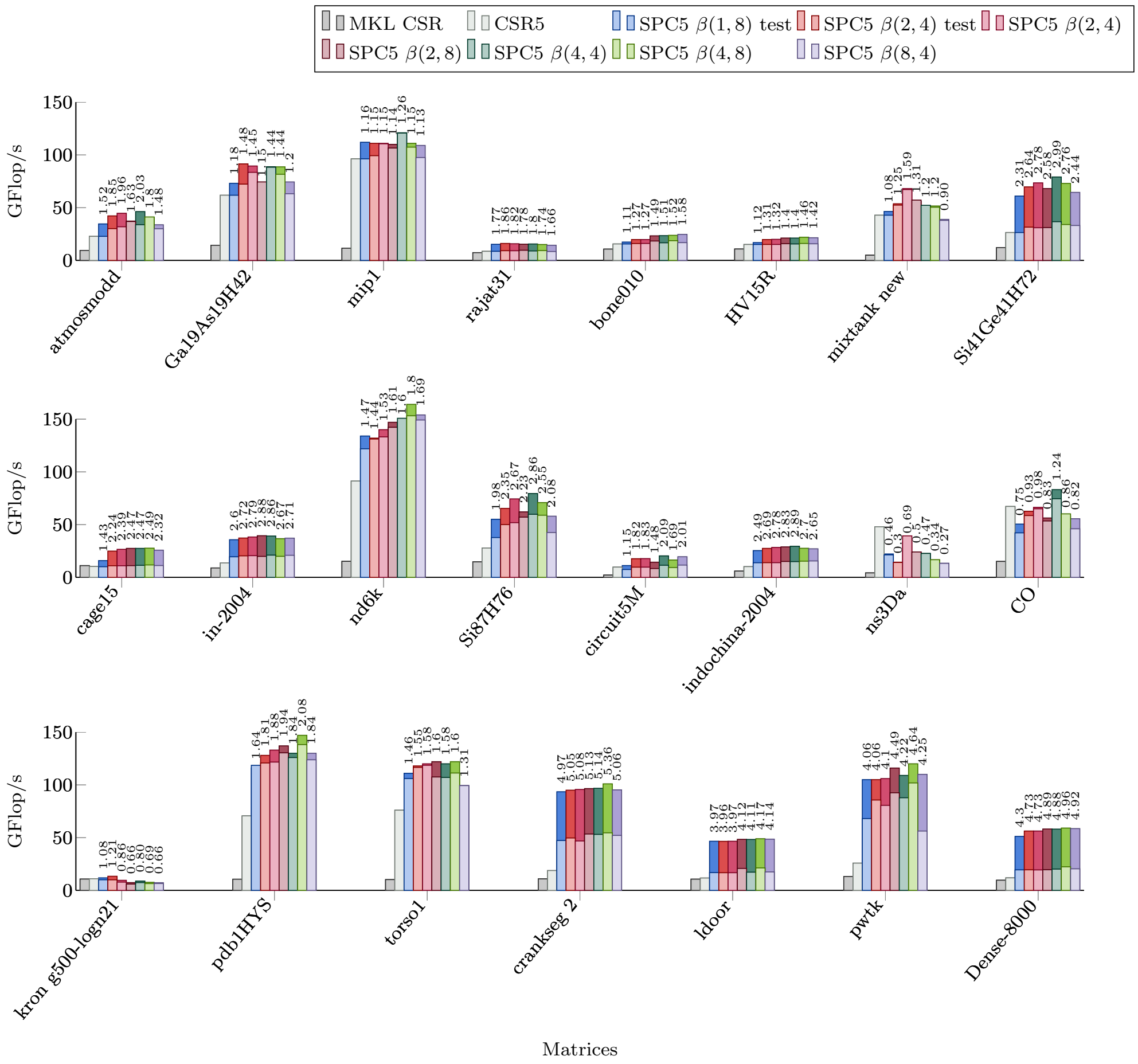

**Figure 4 Performance in Giga Flop per second in double precision for the parallel implementations of MKL CSR, the CSR5 and our SPC5 kernels, all using 52 threads.** Each bar shows the performance without NUMA optimization (light) and with NUMA optimization (dark). Speedup of SPC5 with NUMA optimizations against the better of MKL CSR and CSR5 is shown above the bars.

## Parallel SpMV performance

Figure 4 shows the performance of parallel versions of all investigated kernels using 52 cores for the matrices from the Set-A. The parallel version of MKL CSR is not able to take significant advantage of this number of threads, and it is faster only for *kron g500-logn21* matrix. The CSR5 package is efficient especially when the matrix's structure makes it difficult to achieve high flop-rate. In these cases it has a performance similar or higher than our block-based kernels. All our kernels have very similar performances. By comparing Fig. 4 with Table 1 we can observe that the NUMA optimization gives noticeable speedup for large matrices (shown as the dark part of bars in Fig. 4. Indeed, when a matrix is allocated on a single memory socket, any access by threads on a different socket is very expensive. This might not be a severe problem if the matrix (or at least the part used by the threads) fits in the L3 cache. Then, only the first access is costly, especially when the data is read-only. On the other hand, if the matrix does not fit in the L3 cache, multiple expensive memory transfers will take place during the computation without any possibility for the CPU to hide them.

## PERFORMANCE PREDICTION AND OPTIMAL KERNEL SELECTION

In the previous section, we compared our method with MKL and CSR5 and showed a significant speed-up for most matrices using any block size. The question that arises, however, is what block size should the user choose to maximize the performance. As it can be seen in Figs. 3 and 4, the performance of individual kernels for different block sizes varies and the best option depends on the matrix.

If the user applies the presented library on many matrices with a similar structure, it is quite likely that he can find an optimal kernel in just several trials. If, however, the user does not have any idea about the structure of the given matrix and no experience with execution times of individual kernels, some decision-making process should be supplied. There are many possible approaches to address this challenging problem. If it should be usable, it needs to be computationally cheap, without conversion of the matrix, yet still able to advise the user about the block size to take.

In the following subsection, we describe several possible approaches towards this problem. The matrices from the Set-A listed in Table 1 are used to fine-tune the prediction method. To assess each of the methods, we introduce new, independent set of matrices labeled Set-B and listed in Table 2.

### Performance polynomial interpolation (sequential)

Figure 5 shows the dependence of kernel performance on an average number of NNZ per block. Each kernel-matrix combination is plotted as a single dot. One can clearly see a correlation between the two quantities, slightly different for each kernel. Polynomial interpolation has been done for each kernel using the matrices from Set-A and its results are shown in Fig. 5.

The Avg.NNZ/blocks numbers can be obtained without converting the matrices into a block-based storage. This can be used to roughly estimate the performance of each individual kernel. It is clear, however, that even within the matrix set used for interpolation, there are matrices with a significantly different performance from the one suggested by the interpolation curve. This difference illustrates the fact that the Avg.NNZ/block metric is a very high-level information that hides all the memory accesses patterns, which can be different depending on the positions of the blocks and their individual structure. If the matrix is small, for example, it can even fit into the cache, which would influence the performance significantly. Much less obvious but equally important is the influence of block distribution on the access patterns of the $x$ array. These effects cannot be determined from simple knowledge of Avg.NNZ/blocks and, therefore, we do not expect very accurate results. On the other hand, as this approach is very cheap, it can be incorporated into our package very easily and can provide some basic guideline.

**Table 2** **Set-B : Matrix set for prediction.**

| Name | Dim | $N_{NNZ}$ | $\frac{N_{NNZ}}{N_{rows}}$ | $\frac{N_{NNZ}}{N_{blocks(1,8)}}$ | $\frac{N_{NNZ}}{N_{blocks(2,4)}}$ | $\frac{N_{NNZ}}{N_{blocks(2,8)}}$ | $\frac{N_{NNZ}}{N_{blocks(4,4)}}$ | $\frac{N_{NNZ}}{N_{blocks(4,8)}}$ | $\frac{N_{NNZ}}{N_{blocks(8,4)}}$ |
|---|---|---|---|---|---|---|---|---|---|
| bundle adj | 513,351 | 20,208,051 | 39 | 5.8 (73%) | 6.8 (85%) | 11 (69%) | 12 (75%) | 21 (66%) | 19 (59%) |
| Cube Coup dt0 | 2,164,760 | 127,206,144 | 58 | 5.9 (74%) | 8 (100%) | 12 (75%) | 16 (100%) | 24 (75%) | 20 (63%) |
| dielFilterV2real | 1,157,456 | 48,538,952 | 41 | 2.6 (33%) | 2.6 (33%) | 3.6 (23%) | 3.6 (23%) | 5.1 (16%) | 4.9 (15%) |
| Emilia 923 | 923,136 | 41,005,206 | 44 | 4.1 (51%) | 5 (63%) | 7 (44%) | 7.5 (47%) | 11 (34%) | 11 (34%) |
| FullChip | 2,987,012 | 26,621,990 | 8 | 2 (25%) | 2.4 (30%) | 2.9 (18%) | 3.3 (21%) | 4.2 (13%) | 4.2 (13%) |
| Hook 1498 | 1,498,023 | 60,917,445 | 40 | 4.1 (51%) | 5.1 (64%) | 6.9 (43%) | 7.7 (48%) | 11 (34%) | 11 (34%) |
| RM07R | 381,689 | 37,464,962 | 98 | 4.9 (61%) | 4.7 (59%) | 8.3 (52%) | 7.6 (48%) | 13 (41%) | 12 (38%) |
| Serena | 1,391,349 | 64,531,701 | 46 | 4.1 (51%) | 5.1 (64%) | 7 (44%) | 7.6 (48%) | 11 (34%) | 11 (34%) |
| spal 004 | 10,203 × 321,696 | 46,168,124 | 4,524 | 6 (75%) | 4 (50%) | 7.3 (46%) | 4.3 (27%) | 8.1 (25%) | 4.4 (14%) |
| TSOPF RS b2383 c1 | 38,120 | 16,171,169 | 424 | 7.6 (95%) | 7.8 (98%) | 15 (94%) | 15 (94%) | 30 (94%) | 29 (91%) |
| wikipedia-20060925 | 2,983,494 | 37,269,096 | 12 | 1.1 (14%) | 1.1 (14%) | 1.1 (7%) | 1.1 (7%) | 1.1 (3%) | 1.1 (3%) |

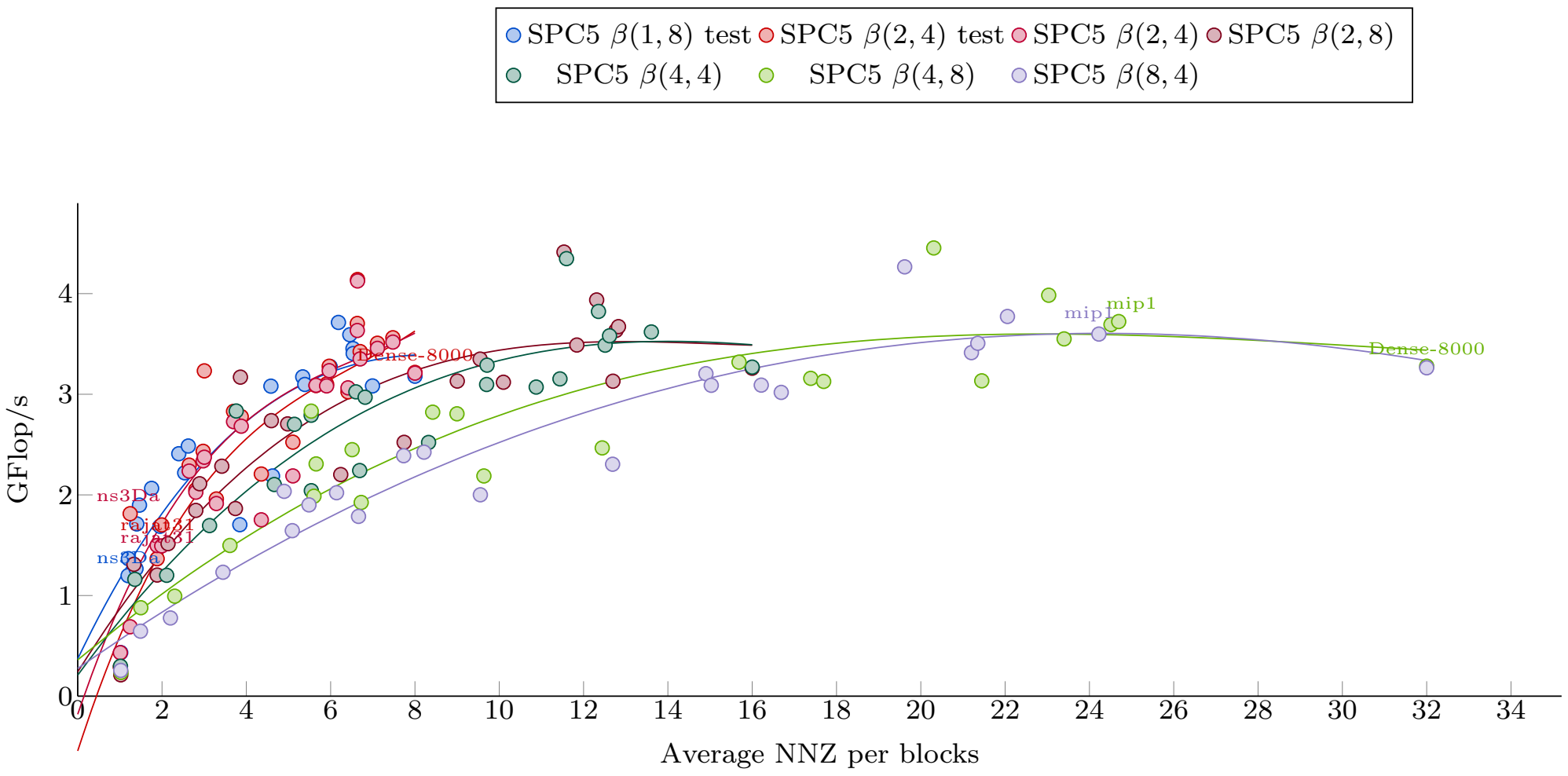

**Figure 5** **Polynomial interpolation of the performance in Giga-Flop per second vs. the average number of values per blocks.** The dots are the matrices from Set-1 used to obtain the polynomial weights.
Full-size DOI: 10.7717/peerjcs.151/fig-5

To select a kernel, we perform the following steps. From a given matrix, we first compute the Avg.NNZ/blocks for the sizes that correspond to our formats. Then, we use the estimation formulas plotted in Fig. 5 to have an estimation of the performance for each kernel. Finally, we select the kernel with the highest estimated performance as the one to use.

We can assess a quality of the prediction system using data shown in Table 3. For each matrix, we provide the objectively best kernel and its speed, the recommended kernel and its estimated and real speed and, finally, the difference between the best and recommended kernels speed.

If the selected kernel is the optimal one, then *Best kernel* has the same value as *Selected kernel* and *Speed difference* is zero. We see that the method selects the best kernels or kernels that have very close performance in most cases, even though the performance estimations as such are not always completely accurate. In other words, the values in columns *Best kernel speed* and *Selected kernel predicted speed* are in most cases quite similar, but there are general differences between values in columns *Selected kernel predicted speed* and *Real speed of selected kernel*.

For some matrices, all kernels have very similar performance and thus, even if the prediction is not accurate and a random kernel is selected, its performance is actually not far from the optimal one. For other matrices, it seems that if they have a special structure that makes the performance far from the one given by the interpolation curves, this is the case for all kernels and so the kernel recommendation is finally good. Of course, one could design a specific sparsity pattern in order to make the prediction fail, but it seems that this simple and cheap prediction system works reasonably well for real-world matrices.

## Parallel performance estimation

Similarly, as we have done for the sequential kernels in the last section, we attempt to estimate the performance of parallel kernels in order to advise the user about the best block size to choose for a given matrix before the matrix is converted into a block-based format. The situation is, however, more complicated since the performance of the kernels will depend on another parameter: the number of threads, which will be used. Thus, we perform a non-linear 2D regression of performance based on two parameters: the number of threads and the average values per block. We use performance results obtained for matrices from Set-A using 1, 4, 16, 32 and 52 threads for each kernel.

The results are then used to estimate (by interpolation) the performance of individual kernels for a given setup. The results of interpolation used on both Set-A and independent Set-B are shown in Fig. 6. We show in Fig. 6A when this strategy leads to an optimal kernel selection. In Fig. 6B, we show what is the real performance difference between the kernel advised and the one that is actually the fastest, and in Fig. 6C, we show the prediction difference for the selected kernel/block size. We observe that the approach does not select the optimal kernels in most cases, but we can also see that the performance provided by the selected kernels are close to the optimal ones (less than 10 percent difference in most cases). Similarly, as for the polynomial interpolation in sequential, this is true despite the performance estimation not being accurate.

**Table 3 Performance estimation and kernel selection for matrices from Set-A and Set-B.** The selection is done by having a polynomial interpolation based on the results of matrices from Set-A. *Speed difference* shows the performance difference between the selected kernel and the best one (if it is 0, the algorithm selected the optimal kernel—Best kernel and Selected kernel are the same).

| Matrices | Best kernel | Best kernel speed | Selected kernel | Selected kernel predicted speed | Real speed of selected kernel | Speed difference |
|---|---|---|---|---|---|---|
| atmosmodd | $\beta(4,4)$ | 2.10 | $\beta(4,4)$ | 2.25 | 2.10 | 0.00% |
| Ga19As19H42 | $\beta(4,4)$ | 3.02 | $\beta(4,4)$ | 2.78 | 3.02 | 0.00% |
| mip1 | $\beta(4,8)$ | 3.69 | $\beta(8,4)$ | 3.605 | 3.60 | 2.56% |
| rajat31 | $\beta(2,4)$ test | 1.50 | $\beta(2,4)$ | 1.63 | 1.27 | 15.42% |
| bone010 | $\beta(4,8)$ | 3.16 | $\beta(4,8)$ | 3.48 | 3.16 | 0.00% |
| HV15R | $\beta(4,8)$ | 3.13 | $\beta(4,8)$ | 3.49 | 3.13 | 0.00% |
| mixtank new | $\beta(2,4)$ | 3.23 | $\beta(2,4)$ | 2.31 | 3.23 | 0.00% |
| Si41Ge41H72 | $\beta(4,4)$ | 2.97 | $\beta(4,4)$ | 2.83 | 2.97 | 0.00% |
| cage15 | $\beta(2,4)$ | 1.70 | $\beta(4,4)$ | 1.71 | 1.70 | 0.38% |
| in-2004 | $\beta(4,4)$ | 2.24 | $\beta(2,8)$ | 2.91 | 2.20 | 1.79% |
| nd6k | $\beta(4,8)$ | 3.98 | $\beta(4,8)$ | 3.59 | 3.98 | 0.00% |
| Si87H76 | $\beta(4,4)$ | 2.79 | $\beta(4,4)$ | 2.51 | 2.79 | 0.00% |
| circuit5M | $\beta(4,4)$ | 2.04 | $\beta(4,4)$ | 2.51 | 2.04 | 0.00% |
| indochina-2004 | $\beta(2,4)$ | 2.52 | $\beta(2,8)$ | 3.19 | 2.52 | 0.06% |
| ns3Da | $\beta(2,4)$ | 1.81 | $\beta(1,8)$ test | 1.30 | 1.20 | 33.94% |
| CO | $\beta(4,4)$ | 2.70 | $\beta(4,4)$ | 2.40 | 2.70 | 0.00% |
| kron g500-logn21 | $\beta(1,8)$ test | 0.43 | $\beta(1,8)$ test | 1.18 | 0.43 | 0.00% |
| pdb1HYS | $\beta(4,8)$ | 4.45 | $\beta(4,8)$ | 3.57 | 4.45 | 0.00% |
| torso1 | $\beta(4,8)$ | 3.72 | $\beta(4,8)$ | 3.58 | 3.72 | 0.00% |
| Crankseg 2 | $\beta(2,8)$ | 3.35 | $\beta(2,8)$ | 3.40 | 3.35 | 0.00% |
| ldoor | $\beta(4,8)$ | 3.13 | $\beta(4,8)$ | 3.59 | 3.13 | 0.00% |
| pwtk | $\beta(4,8)$ | 3.55 | $\beta(4,8)$ | 3.59 | 3.55 | 0.00% |
| Dense-8000 | $\beta(4,8)$ | 3.28 | $\beta(2,4)$ test | 3.62 | 3.21 | 2.11% |
| bundle_adj | $\beta(4,8)$ | 1.78 | $\beta(4,8)$ | 3.58 | 1.78 | 0.00% |
| Cube_Coup_dt0 | $\beta(2,8)$ | 1.67 | $\beta(2,4)$ test | 3.62 | 1.67 | 0.10% |
| dielFilterV2real | $\beta(2,8)$ | 2.01 | $\beta(1,8)$ | 2.11 | 1.23 | 38.35% |
| Emilia_923 | $\beta(2,4)$ | 2.73 | $\beta(2,8)$ | 3.06 | 2.71 | 0.54% |
| FullChip | $\beta(4,4)$ | 0.89 | $\beta(2,4)$ | 1.96 | 0.84 | 5.77% |
| Hook_1498 | $\beta(2,4)$ | 2.63 | $\beta(2,8)$ | 3.06 | 2.60 | 0.76% |
| RM07R | $\beta(2,8)$ | 1.70 | $\beta(2,8)$ | 3.26 | 1.70 | 0.00% |
| Serena | $\beta(4,8)$ | 1.16 | $\beta(2,8)$ | 3.06 | 1.13 | 2.78% |
| spal_004 | $\beta(4,8)$ | 1.78 | $\beta(1,8)$ | 3.21 | 1.63 | 8.02% |
| TSOPF_RS_b2383_c1 | $\beta(4,4)$ | 2.19 | $\beta(2,4)$ test | 3.56 | 1.97 | 11.16% |
| wikipedia-20060925 | $\beta(1,8)$ | 0.44 | $\beta(1,8)$ | 1.20 | 0.44 | 0.00% |

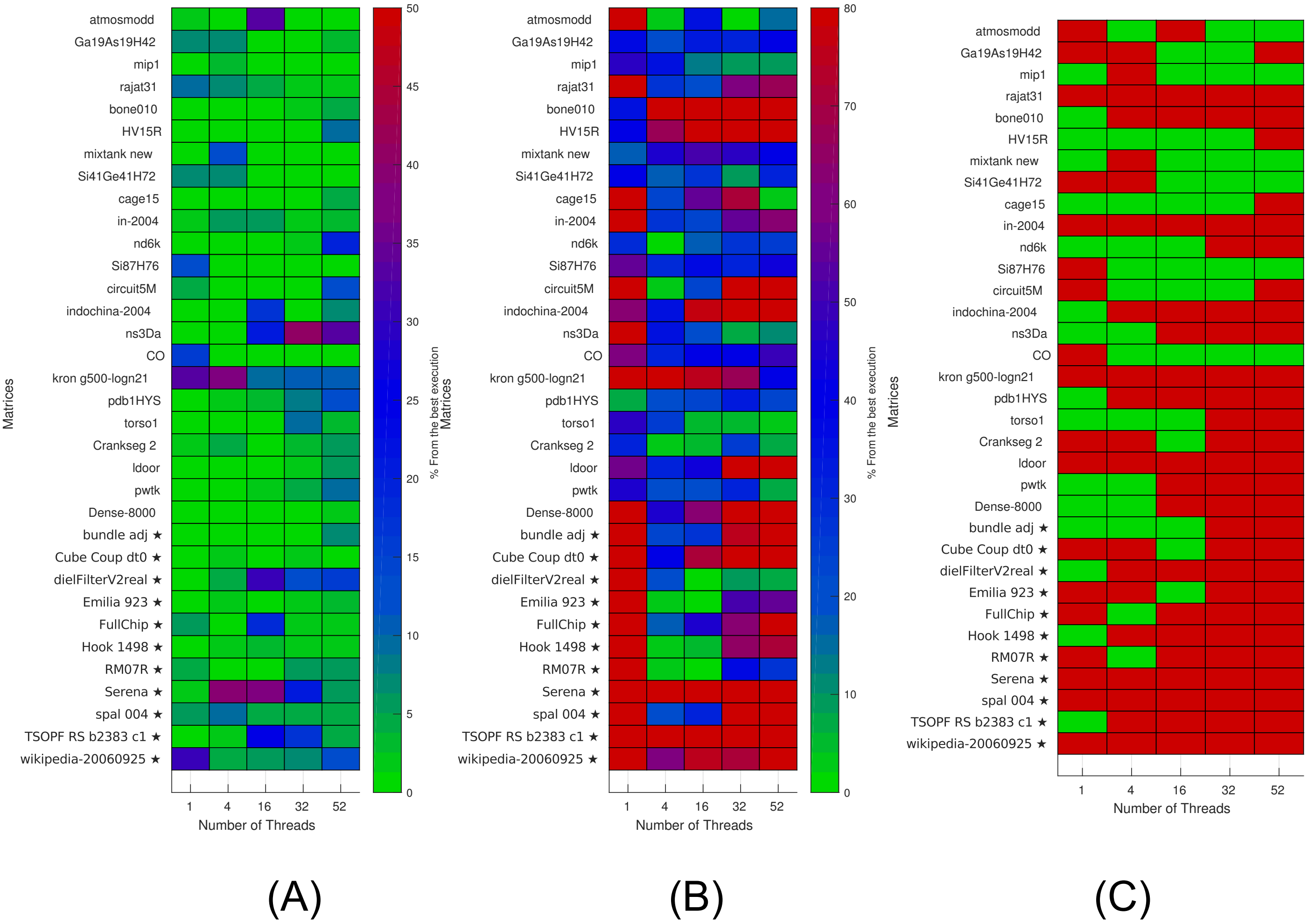

**Figure 6** **Selection of kernels for matrices of Set-A and Set-B by non-linear interpolation.** Matrices from Set-A are used to find the interpolation coefficients. Interpolation is then applied for both Set-A and independent Set-B (matrices labeled with ⋆). (A) Selection of the optimal kernels. Green means success, red failure. (B) Performance difference between the selected kernels and the best ones. (C) Difference between the real and estimated performance of the selected kernels.

Full-size DOI: 10.7717/peerjcs.151/fig-6

## CONCLUSIONS

In this paper, we described new block-based sparse matrix formats without padding. Their corresponding SpMV kernels are fully vectorized, thanks to the *vexpand* AVX-512 instruction. We implemented optimized routines for certain block sizes, and the corresponding source code is available in the SPC5 open-source package. The performance of our kernels shows a considerable increase in speed compared to two reference libraries. Furthermore, our strategy is likely to benefit from the expected increase of the SIMD vector length in the future. The conversion of a matrix, originally in the CSR format, to one of our block-based formats is usually two times the duration of a SpMV computation. Therefore, even if the matrix is not directly assembled in our format (which would be the ideal case, but it might be cumbersome since it usually requires changes in the user code), conversion can be justified, e.g., in the case of iterative methods, where many SpMVs are performed. Moreover, for the $\beta(1,8)$ format, the array of values is unchanged, and only an extra array to store the blocks' masks is needed, compared to the CSR storage. Since, at the same time, the length of *colidx* array is reduced, block storage leads to smaller memory requirements. We have also shown, that when used in parallel, our algorithm can further benefit significantly from reducing the NUMA effect by splitting the arrays, particularly for large matrices. This approach, however, has some drawbacks. It splits the arrays for a given thread configuration so that later, modifications or accesses to the entries become more complicated or even more expensive within some codes. For this reason, we provide two variants of kernels, letting the user to choose whether to optimize for NUMA effects or not, and we show the performance of both.

The kernels provided by our library are usually significantly faster than the competing libraries. To use the library to its full extent, however, it is important for the user to predict which format (block size) is the most appropriate for a given matrix. This is why we also provide techniques to find the optimal kernels using simple interpolation to be used in cases, when the user is unable to choose based on his own knowledge or experience. The performance estimate is usually not completely accurate, but the selected format/kernel is very close to the optimal one, which makes it a reliable method.

In the future, we would like to develop more sophisticated best kernel prediction methods with multiple inputs, such as statistics on the blocks and some hardware properties, the cache size, the memory bandwidth, etc. However, this will require having access to various Skylake-based hardware and running large benchmark suites. In addition, we would like to find a way to incorporate such methods in our package without increasing the dependency footprint. At the optimization level, we would like, among other things, to assess the benefit and cost of duplicating the $x$ vector on every memory node within the parallel computation.

## ADDITIONAL INFORMATION AND DECLARATIONS

### Funding

The authors received no funding for this work.

### Competing Interests

Berenger Bramas is a researcher at MPCDF. Pavel Kus is a researcher at MPCDF.

### Author Contributions

- Bérenger Bramas and Pavel Kus conceived and designed the experiments, performed the experiments, analyzed the data, contributed reagents/materials/analysis tools, prepared figures and/or tables, performed the computation work, approved the final draft.

### Data Availability

The following information was supplied regarding data availability:

GitHub: https://gitlab.mpcdf.mpg.de/bbramas/spc5